\documentclass[12pt,a4paper]{article}
\usepackage[english, USenglish]{babel}
\usepackage{amsmath}
\usepackage{amsfonts}
\usepackage[bookmarksnumbered=true]{hyperref}
\usepackage{a4}
\newcommand{\starco}{\stackrel{*}{,}}
\newcommand{\inv}[1]{\frac{1}{#1}}
\newcommand{\nuu}{{\phantom{1}\nu}}

\newcommand{\rhoo}{{\phantom{1}\rho}}

\author{\\[-0.3cm]\Large M.~Attems\thanks{{\tt mattems@hep.itp.tuwien.ac.at}}~, D.N.~Blaschke\thanks{{\tt blaschke@hep.itp.tuwien.ac.at}}~, S.~Hohenegger\thanks{{\tt hohen@hep.itp.tuwien.ac.at}}~,\\\Large M.~Schweda\thanks{{\tt mschweda@tph.tuwien.ac.at}}~, S.~Stricker\thanks{{\tt stricker@hep.itp.tuwien.ac.at}}}
\title{\bf{Gauge (in)dependence and UV/IR mixing}}
\dateUSenglish
\date{February 22, 2005}
\begin{document}
\begin{titlepage}
\maketitle
\renewcommand{\thefootnote}{\fnsymbol{footnote}}
\begin{center}
\vspace{-0.3cm}Institute for Theoretical Physics, Vienna University of Technology\\Wiedner Hauptstrasse 8-10, A-1040 Vienna, Austria\\[0.5cm]\footnotemark[4]work supported by "Fonds zur F\"orderung der Wissenschaftlichen Forschung" (FWF) under contract P15463-N08\\
\footnotemark[1]\footnotemark[2]\footnotemark[3]\footnotemark[5]work supported by FWF under contract P15015-N08\vspace{0.5cm}
\end{center}
\begin{abstract}
The gauge independence in connection with the UV/IR-mixing is discussed with the help of the non-commutative U(1)-gauge field model proposed by A.~A. Slavnov~\cite{Slavnov} with two different gauges: the covariant gauge fixing defined via a gauge parameter $\alpha$ and the non-standard axial-gauge depending on a fixed gauge direction $n^{\mu}$.
\end{abstract}
\thispagestyle{empty}
\end{titlepage}
%%%%%%%%%%%%%%%%%%%%%%%%%%%%%%%%%%%%%%%%%
\section{Introduction}
%%%%%%%%%%%%%%%%%%%%%%%%%%%%%%%%%%%%%%%%%
In this short letter we investigate the UV/IR mixing and its gauge (in)dependence of a non-commutative U(1) gauge field model with the so-called Slavnov-term in two different gauges. One possibility is the covariant gauge fixing~\cite{Lautrup}, another is the axial gauge~\cite{Kummer,boresch}. Due to the non-commutativity of the Moyal-correspondence together with the star product of two gauge fields \cite{Filk}
\begin{align}\label{star_prod_gauge_fields}
A_\mu(x) \star A_\nu ( x ) = 
e^{\frac{i}{2}\theta^{\mu\nu}\partial^x_\mu\partial^y_\nu} A_\mu(x) A_\nu(y)\Big|_{x = y}\,,
\end{align}
the theory exhibits a non-Abelian structure entailing the usual BRS-quantization \cite{boresch}. The deformation parameter $\theta^{\mu\nu}$ describes the quantized space-time implying the following relation
\begin{align}
\left[ x^\mu \starco x^\nu \right] = i \theta^{\mu\nu},
\end{align}
with the help of the star product where $x^\mu$ are the ordinary commuting coordinates. Thus, through the star product (\ref{star_prod_gauge_fields}) it is now possible to formulate non-commutative quantum field models in terms of ordinary commuting coordinates. The very simple recipe demands the replacement of field products by star products in any action. However, the presence of the $\star$ in the bi-linear part has no effect - therefore the propagators of the field models remain unchanged, whereas the polynomial interaction products of the fields develop additional phases responsible for the so-called UV/IR-mixing. This means that these phases act as a regularization for high internal momenta but produce a new IR- singularity for small external momenta in one loop corrections. These new types of IR-singularities represent a severe obstacle for the renormalization programme at higher order and therefore lead to inconsistencies. The IR-singularities are produced by the so-called UV-finite non-planar one loop graphs in U(N) gauge models and also in scalar field theories. One also has to stress that the usual UV-divergences may be removed by the standard renormalization procedure.

The aim of this letter is devoted to re-investigate a modified non-commutative U(1)-gauge field model proposed by A. Slavnov~\cite{Slavnov}.
A further motivation of the present study is based on the works~\cite{Slavnov,axslavnov}, and an investigation in progress~\cite{paper1} showing that the infrared divergence in the gauge field polarization emerging from non-planar one loop corrections is gauge independent. Explicit calculations carried out in a covariant gauge depending on a gauge parameter $\alpha$ reveal that this infrared divergence is of the form 
\begin{align}
\Pi^{\text{np}}_{\mu\nu}(k) \sim g^2 
\frac{\tilde k_\mu \tilde k_\nu}{ \big({\tilde k}^2\big)^2}\,,
\label{gauge_field_polar}
\end{align}
where $\tilde k^\mu = \theta^{\mu\nu} k_\nu$. The contribution (\ref{gauge_field_polar}) is UV-finite but divergent for small $k$. Additionally, (\ref{gauge_field_polar}) is transverse with respect to $k^\mu$ corresponding the gauge(BRS)-invariance and does not depend on the gauge parameter $\alpha$ (see Sect.~\ref{sect2}).

The model proposed by A. Slavnov~\cite{Slavnov} implies that the gauge field propagator is transverse with respect to $\tilde k^\mu$. In momentum space this means
\begin{align}
\tilde k^\mu \Delta_{\mu\nu}(k) = 0.
\label{transvers}
\end{align}
In order to make the modified U(1) gauge model meaningful, the validity of (\ref{transvers}) must be guaranteed in any gauge fixing procedure implying different gauge field propagators. Therefore, it is the aim of this short contribution to show that (\ref{transvers}) holds for covariant gauges and also for non-standard axial gauges.

The gauge (in)dependence of (\ref{gauge_field_polar}) and (\ref{transvers}) seems to be a fundamental property of the U(1) non-commutative gauge model~\cite{ruiz}. Usually, gauge independence signals physical aspects. But neither quantities, the polarization (\ref{gauge_field_polar}) nor the condition (\ref{transvers}), are physical objects.

Relation (\ref{transvers}) implies
\begin{align}
\Delta_\mu^\rhoo(k) \Pi^{\text{np}}_{\rho\sigma}(k) \Delta^\sigma_\nuu(k) = 0,
\end{align}
leading to a consistent theory at the two loop level.
%%%%%%%%%%%%%%%%%%%%%%%%%%%%%%%%%%%%%%%%%
\section{Non-commutative U(1)-gauge theory at the classical level}\label{sect2}
%%%%%%%%%%%%%%%%%%%%%%%%%%%%%%%%%%%%%%%%%
Corresponding to the work in progress~\cite{paper1} we define the non-commutative U(1)-Maxwell theory in a BRS - invariant manner including also the ghost and anti-ghost fields.

In the sense of Slavnov one has the following classical action for a non-commutative $U(1)$-Maxwell theory for the covariant gauge fixing procedure with the gauge field $A^{\mu}$ \cite{Becchi}:
\begin{align}
\Gamma^{(0)} = \int d^4x
\left[ - \inv 4 F_{\mu\nu} \star F^{\mu\nu} + \frac{\alpha}{2} B \star B +
B \star \partial^\mu A_\mu + \frac{\lambda}{2}\star\theta^{\mu\nu} F_{\mu\nu}
- \bar c\star\partial^\mu D_\mu c \right],
\label{action}
\end{align}
where $F_{\mu\nu}$ is defined by
\begin{align}
F_{\mu\nu} = \partial_\mu A_\mu - \partial_\nu A_\mu - ig [ A_\mu \starco A_\nu],
\label{electro_magnetic_fieldtensor}
\end{align}
and $D_\mu c$ is given through
\begin{align}
D_\mu c = \partial_\mu c - ig [A_\mu \starco c].
\label{total_derivative}
\end{align}
The $B$ field controls the gauge fixing procedure \cite{Lautrup} whereas the multiplier field $\lambda$ of Slavnov implements a new constraint equation responsible for (\ref{transvers}). $\alpha$ is some gauge parameter. $c$ and $\bar c$ are the ghost and anti-ghost fields. The action (\ref{action}) is characterized by the supersymmetric, nilpotent and non-linear BRS transformations~\cite{Piguet}
\begin{align}
& sA_\mu = D_\mu c, && sc=igc \star c, \nonumber\\
& s \bar c = B, && sB = 0, \nonumber \\
& s \lambda = ig [c \starco \lambda ],\nonumber\\
&s^2\phi=0, \hspace{1cm} \text{for}\ \phi=\{A^{\mu},B,c,\bar{c},\lambda\}. 
\label{brs}
\end{align}
Corresponding to the BRS quantization procedure the symmetry content of (\ref{action}) may be characterized by the non-linear Slavnov identity~\cite{Piguet}. Additionally, there are also the gauge fixing condition
\begin{align}
\frac{\delta \Gamma^{(0)}}{\delta B} = \alpha B + \partial^\mu A_\mu=0,
\end{align}
and the constraint equation
\begin{align}
\frac{\delta \Gamma^{(0)}}{\delta \lambda} &= \inv 2 \theta^{\mu\nu} 
F_{\mu\nu} = - \widetilde\partial^\mu A_\mu - \inv 2 ig \theta^{\mu\nu} [A_\mu \starco A_\nu]=0.
\end{align}
In order to calculate the propagator one considers the Legendre-transformation with sources $j_\mu$, $j$ and $j_B$:
\begin{align}\label{action_no_c}
Z_{\text{bi}}^{c} &= \int d^4 x \Big[ - \inv 4 (\partial_\mu A_\nu 
- \partial_\nu A_\mu )^2 + \frac{\alpha}{2} B^2 + B \partial^\mu A_\mu \nonumber\\
&\hspace{1.5cm} - \lambda \widetilde\partial^\mu A_\mu + j_\mu A_\mu + j_B B + j\lambda \Big]=\nonumber\\
&=\Gamma^{(0)}_{\text{bi}}+\int d^4x\left(j_{\mu}A^{\mu}+j_{B}B+j\lambda\right),
\end{align}
where we have suppressed the ghost parts. In order to calculate the propagators of the bosonic sector one has to solve the following system of equations
\begin{align}
\partial^\mu (\partial_\mu A_\nu - \partial_\nu A_\mu) + \widetilde\partial_\mu
\lambda - \partial_\mu B &= - j_\nu, \nonumber\\
- \widetilde\partial^\mu A_\mu &= -j, \nonumber \\
\alpha B + \partial^\mu A_\mu &= - j_B,  
\end{align}
with the solutions
\begin{subequations}
\begin{align}
A_\nu \left[ j_\rho, j , j_B \right] &= - \inv \square \Big[
\left( g_{\nu\rho} - ( 1 - \alpha) \frac{\partial_\nu \partial_\rho}{\square}
\right) - \frac{\widetilde\partial_\nu \widetilde\partial_\rho}{\widetilde\square}\Big] 
j^\rho - \inv \square \partial_\nu j_B + \inv{\widetilde\square} \widetilde\partial_\nu j,\\
\lambda[j^\mu, j] &= - \inv{\widetilde\square} (\widetilde\partial^\mu j_\mu + 
\square j),\\
B[j_\mu] &= \inv \square \partial^\mu j_\mu.
\label{b_solution}
\end{align}
\end{subequations}
For the gauge field propagator this implies
\begin{align}
\Delta^c_{\mu\nu} (k) = \inv {k^2} \left[ g_{\mu\nu} - (1- \alpha) 
\frac{k_\mu k_\nu}{k^2} - \frac{\tilde k_\mu \tilde k_\nu}{\tilde k^2} \right].
\label{gauge_field_propagator}
\end{align}
Due to $\tilde k^\mu k_\mu = 0$ one has the desired result
\begin{align}
\tilde k^\mu \Delta^c_{\mu\nu}(k) = 0,
\end{align}
implied by $\widetilde{\partial}^\mu A_\mu = 0$ for $j=0$.

In order to investigate the (homogeneous) axial gauge one simply replaces $\frac{\alpha}{2}B^2 + B \partial^\mu A_\mu$ in (\ref{b_solution}) with $Bn^\mu A_\mu$ \cite{Kummer,boresch}, where $n^\mu$ is some gauge direction. Therefore, one has
\begin{align}\label{changein}
\frac{\delta \Gamma^{(0)}}{\delta B} = n^\mu A_\mu=0,
\end{align}
in the axial gauge. Thus (\ref{action_no_c}) is changed into
\begin{align}
Z^{c,a}_{bi} = \int d^4x \left[ - \inv 4 (\partial_\mu A_\nu - \partial_\nu A_\mu)^2 + B n^\mu A_\mu 
- \lambda \widetilde\partial^\mu A_\mu + j_\mu A^\mu + j_B B + j\lambda \right],
\label{action_bi}
\end{align}
with the equations of motion
\begin{align}
\partial^\mu (\partial_\mu A_\nu - \partial_\nu A_\mu) + \widetilde\partial_\nu 
\lambda + B n_\nu &= - j_\nu, \nonumber\\
- \widetilde\partial^\mu A_\mu &= -j, \nonumber \\
n^\mu A_\mu &= - j_B.
\label{system}
\end{align}
The solution of (\ref{action_bi}) is given by
\begin{subequations}
\begin{align}
A_\nu\left[ j^\rho, j, j_B \right] &= - \inv \square \Big[
g_{\nu\rho} - \frac{n_\nu \partial_\rho + n_\rho \partial_\nu}{(n\partial)} +
n^2 \frac{\partial_\nu \partial_\rho}{(n \partial)^2} \nonumber\\
&\quad +\frac{n \widetilde\partial}{(n \partial) \widetilde\square}
(\partial_\nu \widetilde\partial_\rho + \partial_\rho \widetilde\partial_\nu)
- \frac{(n\widetilde\partial)^2}{(n \partial)^2 \widetilde\square} \partial_\nu
\partial_\rho
- \frac{\widetilde\partial_\nu \widetilde\partial_\rho}{\widetilde\square} \Big]
j^\rho \nonumber \\
&\quad - \frac{\partial_\nu}{(n\partial)} j_B - 
\frac{(n \widetilde\partial)}{(n \partial)} \frac{\partial_\nu}{\widetilde\square}j
+\frac{\widetilde\partial_\nu}{\widetilde\square} j,\\
\lambda [ j_\mu , j] &= - \inv{\widetilde\square} \left( \widetilde\partial^\mu
j_\mu + \square j - \frac{(n \widetilde\partial) \partial^\mu}{(n \partial)}j_\mu\right),\\
B[j_\mu] &= - \frac{\partial^\mu j_\mu}{(n \partial)}.
\end{align}
\end{subequations}
The corresponding propagator of the gauge field is
\begin{align}
\Delta^a_{\mu\nu}(k) &= \inv {k^2} \Big[ g_{\mu\nu} -
\frac{n_\mu k_\nu + n_\nu k_\mu}{(nk)} + n^2 \frac{k_\mu k_\nu}{(nk)^2} \nonumber\\
&\quad + \frac{(n \tilde k)}{(nk) \tilde k^2} (k_\nu \tilde k_\mu 
+ k_\mu \tilde k_\nu) -\frac{(n \tilde k)^2}{(nk)^2 \tilde k^2} k_\mu k_\nu
- \frac{\tilde k_\mu \tilde k_\nu}{\tilde k^2} \Big].
\label{propagator_horr}
\end{align}
This horrible propagator has the following desired properties
\begin{align}
\tilde k^\mu \Delta^a_{\mu \nu}(k) = n^\mu \Delta^a_{\mu\nu}(k) = 0.
\label{show_off}
\end{align}
Equation (\ref{show_off}) follows from (\ref{system}). The $\tilde k$ independent terms of (\ref{propagator_horr}) are just the usual propagator in the axial gauge~\cite{boresch}.
%%%%%%%%%%%%%%%%%%%%%%%%%%%%%%%%%%%%%%%%%
\section{Pre-conclusion and outlook}
%%%%%%%%%%%%%%%%%%%%%%%%%%%%%%%%%%%%%%%%%
Motivated by the gauge independence of the non-commutative U(1) non-planar vacuum polarization at one loop level, we have shown that the transversality condition of the gauge propagator with respect to $\tilde k^\mu$ is also gauge independent. This result can be understood by the fact that the Slavnov term is gauge invariant.

Unfortunately, however, it might be possible that the axial propagator (\ref{propagator_horr}) produces $n^{\mu}$-dependent IR-singularities in non-planar one loop contributions for the $U(1)$-vacuum polarization. But gauge or BRS-invariance demands 
\begin{align}
k^{\mu}\Pi_{\mu\nu}^{\text{np},a}(k,n)=0,
\end{align}
implying that the only possible (quadratic) IR-divergent term consistent with dimensional analysis is
\begin{align}\label{remaindiv}
\Pi_{\mu\nu}^{\text{np},a}(k,n)\propto\ \frac{\tilde{k}_{\mu}\tilde{k}_{\nu}}{(\tilde{k}^2)^2}.
\end{align}
Equation (\ref{remaindiv}) may be gauge dependent but fulfills
\begin{align}
\Delta^{a}_{\mu\rho}(k,n)\Pi^{\text{np},a,\rho\sigma}(k,n)\Delta^{a}_{\sigma\nu}(k,n)=0.
\end{align}
So again we have gauge (in)dependence of Slavnov's method and no IR-\-singular\-ities are expected to appear at higher loop calculations.

At this point one also has to stress that the explicit one loop calculations strongly depend on the pole-prescription of the spurious terms $(nk)^{-m}$ of the axial gauge field propagator. For $n^2\neq 0$ one has to use the PV-prescription \cite{boresch}. The case $n^2=0$ is more delicate due to the fact that there exists the so-called Leibbrandt-Mandelstam prescription entailing a further dual gauge direction $n^{\star\mu}=(n^0,-\vec{n})$ \cite{boresch}. The existence of $n^{\star}$ usually produces a more complex tensor structure of $\Pi_{\mu\nu}$ which can be non-local and non-transverse \cite{boresch}.
%%%%%%%%%%%%%%%%%%%%%%%%%%%%%%%%%%%%%%%%%
\section{Summary} 
%%%%%%%%%%%%%%%%%%%%%%%%%%%%%%%%%%%%%%%%%
Contrary to the statements made in Slavnov's paper \cite{axslavnov} that the absence of non-integrable IR-singularities is a direct consequence of an appropriate gauge condition, we have shown the gauge (in)dependence of the Slavnov trick in this letter. The non-existence of non-integrable IR-singularities is present in the covariant gauge characterized by a gauge parameter $\alpha$ \cite{Lautrup} and also in the non-standard axial gauge defined with the help of a fixed gauge direction $n^{\mu}$ \cite{boresch}.

%%%%%%%%%%%%%%%%%%%%%%%%%%%%%%%%%%%%%%%%%

\end{document}